\documentclass{emulateapj}   

\usepackage{lineno}
\usepackage{rotating}
\usepackage{color}
\usepackage{hyperref}
\usepackage{breakurl} 

\def\ltsima{$\; \buildrel < \over \sim \;$}
\def\simlt{\lower.5ex\hbox{\ltsima}} 
\def\gtsima{$\; \buildrel > \over \sim \;$}
\def\simgt{\lower.5ex\hbox{\gtsima}} 
\def\arcsec{\hbox{$^{\prime\prime}$}}
\def\deg{\hbox{$^\circ$}}
\def\phflux{photons cm$^{-2}$ s$^{-1}$}

\def\gray{$\gamma$-ray}
\def\grays{$\gamma$-rays}
\def\dt{\hbox{$\Delta$$t$}}
\def\dtr{\hbox{$\Delta$$t_{\rm r}$}}
\def\dtg{\hbox{$\Delta$$t_{\gamma}$}}

\def\p0{$\pi^{\rm 0}$}

\def\r95{$r_{\rm 95}$}

\def\Fermi{\textit{Fermi}}

\slugcomment{ApJL, accepted for publication}

\shorttitle{Gravitational Lens Delayed Gamma-ray Flares in B0218+357}
\shortauthors{Cheung et al.}

\begin{document}

\title{\Fermi-LAT Detection of Gravitational Lens Delayed Gamma-ray Flares from 
Blazar B0218+357}

\author{
C.~C.~Cheung\altaffilmark{1,2}, 
S.~Larsson\altaffilmark{3,4,5,6,7},
J.~D.~Scargle\altaffilmark{8,9},
M.~A.~Amin\altaffilmark{10}, 
R.~D.~Blandford\altaffilmark{11},
D.~Bulmash\altaffilmark{12},
J.~Chiang\altaffilmark{11},
S.~Ciprini\altaffilmark{13,14},
R.~H.~D.~Corbet\altaffilmark{15,16},
E.~E.~Falco\altaffilmark{17},
P.~J.~Marshall\altaffilmark{18,11},
D.~L.~Wood\altaffilmark{19},
M.~Ajello\altaffilmark{20}, 
D.~Bastieri\altaffilmark{21,22}, 
A.~Chekhtman\altaffilmark{23}, 
F.~D'Ammando\altaffilmark{24}, 
M.~Giroletti\altaffilmark{24}, 
J.~E.~Grove\altaffilmark{1}, 
B.~Lott\altaffilmark{25}, 
R.~Ojha\altaffilmark{26}, 
M.~Orienti\altaffilmark{24}, 
J.~S.~Perkins\altaffilmark{26}, 
M.~Razzano\altaffilmark{27,28}, 
A.~W.~Smith\altaffilmark{29}, 
D.~J.~Thompson\altaffilmark{26}, 
K.~S.~Wood\altaffilmark{1}
}
\altaffiltext{1}{Space Science Division, Naval Research Laboratory, Washington, DC 20375-5352, USA}
\altaffiltext{2}{email: Teddy.Cheung@nrl.navy.mil}
\altaffiltext{3}{Department of Physics, Stockholm University, AlbaNova, SE-106 91 Stockholm, Sweden}
\altaffiltext{4}{The Oskar Klein Centre for Cosmoparticle Physics, AlbaNova, SE-106 91 Stockholm, Sweden}
\altaffiltext{5}{Department of Astronomy, Stockholm University, SE-106 91 Stockholm, Sweden}
\altaffiltext{6}{Supported by the Royal Swedish Academy Crafoord Foundation}
\altaffiltext{7}{email: stefan@astro.su.se}
\altaffiltext{8}{Space Sciences Division, NASA Ames Research Center, Moffett Field, CA 94035-1000, USA}
\altaffiltext{9}{email: Jeffrey.D.Scargle@nasa.gov}
\altaffiltext{10}{Kavli Institute for Cosmology and Institute of Astronomy, University of Cambridge, Madingley Road, Cambridge CB3 0HA, UK}
\altaffiltext{11}{W. W. Hansen Experimental Physics Laboratory, Kavli Institute for Particle Astrophysics and Cosmology, Department of Physics and SLAC National Accelerator Laboratory, Stanford University, Stanford, CA 94305, USA}
\altaffiltext{12}{Department of Physics, Stanford University, Stanford, CA 94305, USA}
\altaffiltext{13}{Agenzia Spaziale Italiana (ASI) Science Data Center, I-00133 Roma, Italy}
\altaffiltext{14}{Istituto Nazionale di Astrofisica - Osservatorio Astronomico di Roma, I-00040 Monte Porzio Catone (Roma), Italy}
\altaffiltext{15}{Center for Research and Exploration in Space Science and Technology (CRESST) and NASA Goddard Space Flight Center, Greenbelt, MD 20771, USA}
\altaffiltext{16}{University of Maryland Baltimore County, Baltimore, MD 21250, USA}
\altaffiltext{17}{Harvard-Smithsonian Center for Astrophysics, Cambridge, MA 02138, USA}
\altaffiltext{18}{Department of Physics (Astrophyics), Oxford University, Oxford OX1 3RH, UK}
\altaffiltext{19}{Praxis Inc., Alexandria, VA 22303, resident at Naval Research Laboratory, Washington, DC 20375, USA}
\altaffiltext{20}{Space Sciences Laboratory, 7 Gauss Way, University of California, Berkeley, CA 94720-7450, USA}
\altaffiltext{21}{Istituto Nazionale di Fisica Nucleare, Sezione di Padova, I-35131 Padova, Italy}
\altaffiltext{22}{Dipartimento di Fisica e Astronomia ``G. Galilei'', Universit\`a di Padova, I-35131 Padova, Italy}
\altaffiltext{23}{Center for Earth Observing and Space Research, College of Science, George Mason University, Fairfax, VA 22030, resident at Naval Research Laboratory, Washington, DC 20375, USA}
\altaffiltext{24}{INAF Istituto di Radioastronomia, 40129 Bologna, Italy}
\altaffiltext{25}{Centre d'\'Etudes Nucl\'eaires de Bordeaux Gradignan, IN2P3/CNRS, Universit\'e Bordeaux 1, BP120, F-33175 Gradignan Cedex, France}
\altaffiltext{26}{NASA Goddard Space Flight Center, Greenbelt, MD 20771, USA}
\altaffiltext{27}{Istituto Nazionale di Fisica Nucleare, Sezione di Pisa, I-56127 Pisa, Italy}
\altaffiltext{28}{Funded by contract FIRB-2012-RBFR12PM1F from the Italian Ministry of Education, University and Research (MIUR)}
\altaffiltext{29}{Department of Physics and Astronomy, University of Utah, Salt Lake City, UT 84112, USA}

\begin{abstract}

Using data from the \Fermi\ Large Area Telescope (LAT), we report the first 
clear \gray\ measurement of a delay between flares from the gravitationally 
lensed images of a blazar. The delay was detected in B0218+357, a known 
double-image lensed system, during a period of enhanced \gray\ activity with 
peak fluxes consistently observed to reach $>20-50\times$ its previous average 
flux. An auto-correlation function analysis identified a delay in the \gray\ 
data of $11.46\pm0.16$ days ($1\sigma$) that is $\sim1$ day greater than
previous radio measurements. Considering that it is beyond the capabilities of 
the LAT to spatially resolve the two images, we nevertheless decomposed 
individual sequences of superposing \gray\ flares/delayed emissions. In three 
such $\sim8-10$ day-long sequences within a $\sim4$-month span, considering 
confusion due to overlapping flaring emission and flux measurement 
uncertainties, we found flux ratios consistent with $\sim1$, thus systematically 
smaller than those from radio observations. During the first, best-defined 
flare, the delayed emission was detailed with a \Fermi\ pointing, and we 
observed flux doubling timescales of $\sim3-6$ hrs implying as well extremely 
compact \gray\ emitting regions.

\end{abstract}
 

\section{Introduction\label{section-intro}}

B0218+357 was discovered with the NRAO 140-ft telescope in its strong source 
survey \citep[S3\,0218+35;][]{pau72}. Later radio imaging revealed it to be a 
gravitationally lensed blazar with the smallest separation double-image known 
(335 milli-arcseconds), and an Einstein ring with a similar angular diameter 
\citep{ode92,pat93}. The lens galaxy is at redshift $z=0.6847$ \citep{bro93}, 
and the blazar was later securely measured at $z=0.944\pm0.002$ \citep{coh03}.

Shortly after the lens discovery, \citet{cor96} measured a time delay 
\citep{ref64} $\dtr=12\pm3$ days ($1\sigma$ quoted throughout unless otherwise 
specified) at radio wavelengths, using the VLA to spatially separate and monitor 
the polarization variability in its leading brighter A (western) and fainter B 
(eastern) images. Later independent (but contemporaneous) dual-frequency VLA 
observations further refined the delay, $\dtr=10.5\pm0.2$ \citep{big99} and 
$10.1\pm0.8$ days \citep{coh00}. Interestingly, \citet{eul11} analyzed the 
latter's measurements and found two possible delays, $\dtr=9.9^{+4.0}_{-0.9}$ or 
$11.8\pm2.3$ days. Although these delays span a narrow range, $\dtr\sim10-12$ 
days, because of the differing assumptions and analysis techniques employed in 
these works, there remains some debate on how to best derive their 
uncertainties.

\begin{figure*}
\epsscale{1.0}
\plotone{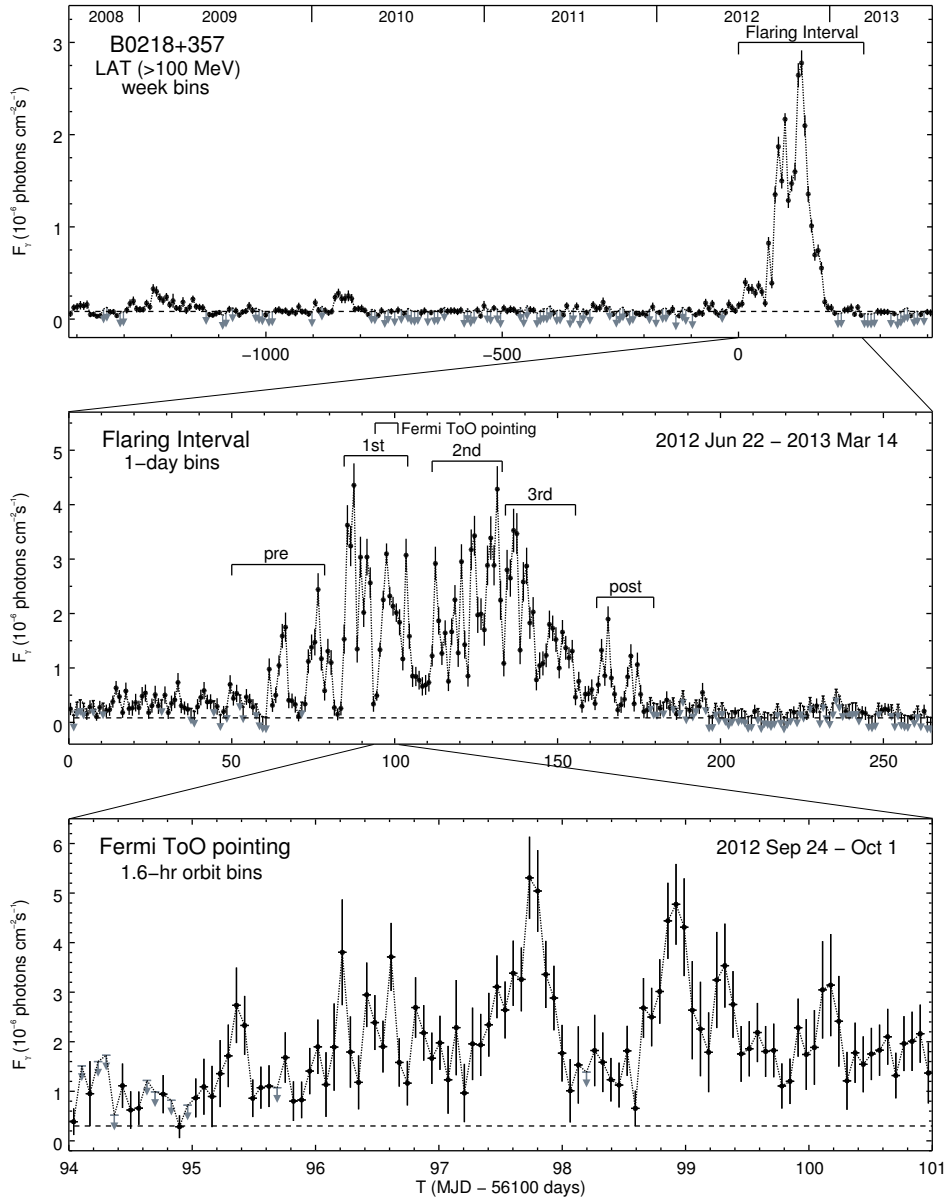}
\figcaption[f1.eps]{\label{figure-1}
LAT lightcurves over wide dynamic range: 1-week bins over the first 5-years of
the \Fermi\ mission (top), 1-day bins for a 265-day flaring interval (middle),
and 1.6-hr orbit bins during the 7-day \Fermi\ ToO (bottom). The pre, post, and
three active episodes outlined in the middle panel are further detailed in
Figure~\ref{figure-2}. Throughout, flux points (plotted with $1\sigma$ errors
when $TS\geq4$ in the bin) and arrows indicating $2\sigma$ upper limits (when
$TS<4$) are connected by dotted lines. Horizontal dashed lines indicate the
3.9-year average flux prior to the flaring interval (top) and the baseline
flux during the flaring interval, $F_{\gamma}=0.3\times10^{-6}$ \phflux\
(middle, bottom).} \end{figure*}

B0218+357 is also a \gray\ source detected by the \Fermi\ Large Area Telescope 
\citep[LAT;][]{atw09} with an average flux\footnote{LAT \gray\ fluxes are 
reported at $E>100$ MeV throughout.} $F_{\gamma}=(1.00\pm0.07)\times10^{-7}$ 
\phflux\ over its first 2-years of observations 
\citep[2FGL~J0221.0+3555;][]{2fgl}. Its steep spectrum at $>100$ MeV energies 
(photon index, $\Gamma=2.28\pm0.04$) and overall spectral energy distribution 
are typical of an otherwise normal \gray\ emitting flat-spectrum radio quasar 
\citep[e.g.,][]{ghi10}. While \gray\ data lack the necessary spatial resolution 
to separate lensed images, such blazars display their most dramatic variability 
in \grays, and the LAT's all-sky monitoring could give it a distinct advantage 
over lower-frequency imaging observations in parameterizing lensed systems. 
Indeed, \citet{atw07} proposed prior to \Fermi's launch that the LAT could 
detect delayed emission from such gravitationally lensed blazars using 
integrated lightcurves for sufficiently bright \gray\ flares. B0218+357 was 
found to be variable in the early LAT observations, though only modestly so 
\citep[][see Figure~\ref{figure-1}]{lbasvar}.

Bright \gray\ flaring from B0218+357 was detected with the LAT beginning late 
2012 August \citep{cip12}, and a delayed flare was tentatively identified 
$\sim10$ days later \citep{gir12}, consistent with the radio delay measurements. 
The blazar then displayed even brighter, more sustained flaring activity 
beginning September 14, thus prompting a \Fermi\ Target-of-Opportunity (ToO) 
pointed observation \citep{che12} that traced the anticipated delayed emission 
in detail. Two main additional flaring events were subsequently observed in as 
many months (see Figure~\ref{figure-1} for an overview). We discuss the \gray\ 
temporal and spectral properties of B0218+357 together with the derived time 
lag, flare timescales, and observed flux ratios of the A/B images.

\section{LAT Observations and Analysis}

The \Fermi-LAT operates in a default sky-survey mode, and over every two 
$\sim$1.6-hr spacecraft orbits, provides observations covering the entire sky. 
We used LAT observations with the {\tt P7SOURCE$\_$V6} instrument response 
functions, selecting 100 MeV -- 100 GeV events with a region of interest (ROI) 
of radius = 15\deg\ centered at the B0218+357 radio position, 
${\rm~R.A.}=35\deg.27279$, ${\rm~Decl.}=35\deg.93715$ \citep[J2000;][]{pat92}. 
The maximum zenith angle of 100\deg\ was set to minimize the contamination from 
Earth limb photons as well as the appropriate {\tt gtmktime} filter ($\#$3) 
following the FSSC 
recommendations\footnote{\burl{http://fermi.gsfc.nasa.gov/ssc/data/analysis/documentation/Cicerone/Cicerone_Likelihood/Exposure.html}} 
for the combination of sky-survey and pointed observations. The {\tt gtlike} 
likelihood in the $Fermi$ Science tools (version {\tt v9r27p1}) was used for the 
spectral analysis, assuming throughout a single power-law model for B0218+357 
over the selected energy range (as in the 2FGL catalog). The background model 
included all 2FGL sources within the ROI as well as the Galactic ({\tt 
gal$\_$2yearp7v6$\_$v0.fits}) and isotropic ({\tt iso$\_$p7v6source.txt}) 
diffuse components.

In generating each lightcurve, the isotropic normalization was left free to vary 
in each time-bin while the two known variable 2FGL sources within a $5\deg$ ROI 
and the Galactic normalization were initially fitted over each full interval, 
then fixed at the average fitted values in the shorter time-bins. As a 
convenient reference point, we define $T={\rm~MJD} - 56100$ days (i.e., $T=0$ 
was 2012 June 22), the time when \gray\ flaring became obvious. Integrating 1417 
days ($\sim3.9$ years) of LAT observations prior to this date gave an average 
$F_{\gamma}=(0.83\pm0.05)\times10^{-7}$ \phflux, with $\Gamma=2.30\pm0.03$, 
consistent with the 2FGL value. For context, we generated a 1-week binned 
lightcurve for 5-years of data (2008 August 5 - 2013 August 6; 
Figure~\ref{figure-1}, top) assuming a fixed $\Gamma=2.3$. Besides the modest 
source activity in early 2009 and 2010, the pronounced flaring beginning in 
mid-2012 lasting for $\sim200$ days is apparent; thereafter, the source quieted 
again to earlier levels.

\begin{figure*}
\hspace*{-1.25cm}
\includegraphics[scale=.72]{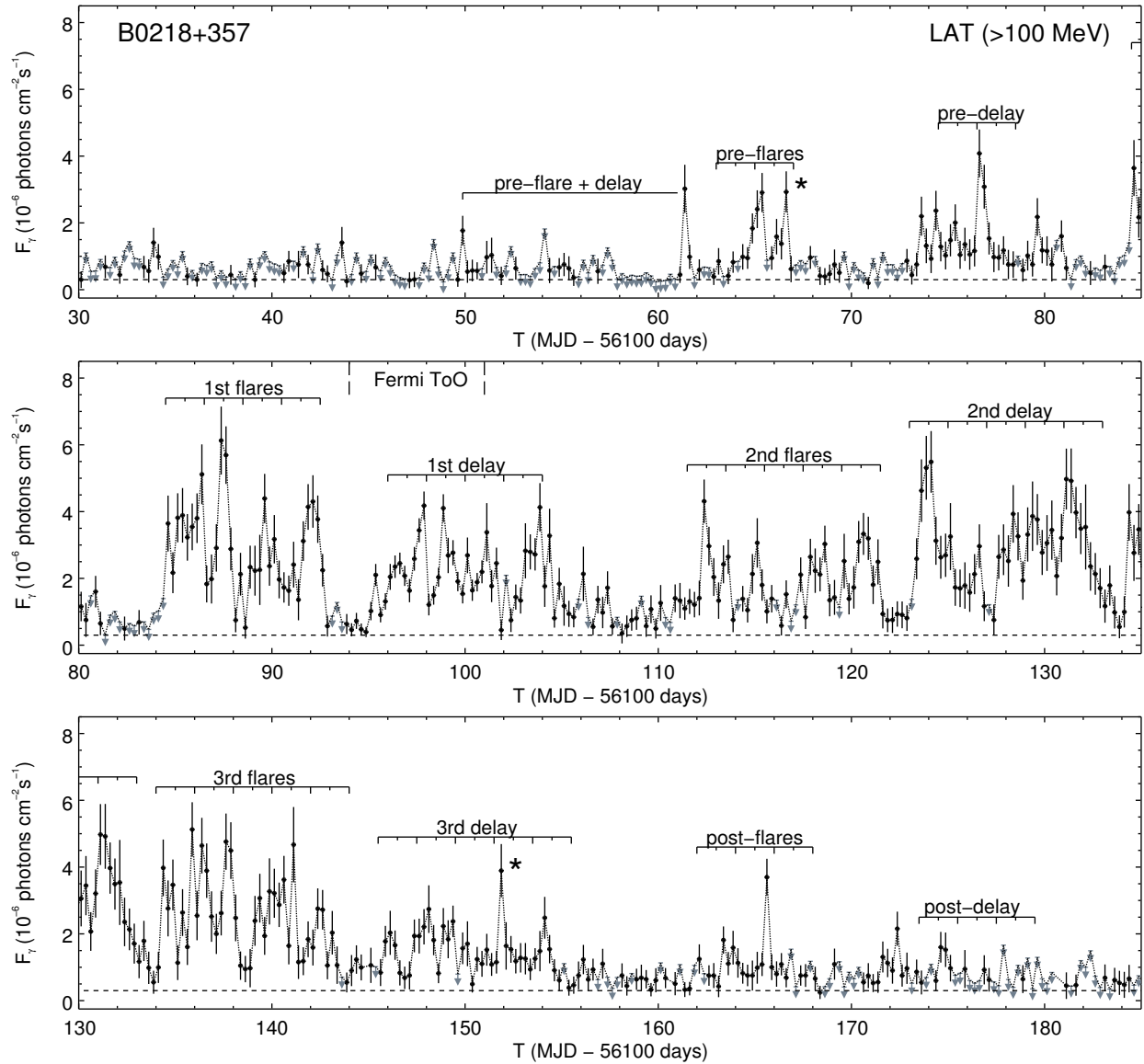}
\figcaption[f2.eps]{\label{figure-2}
\small LAT lightcurve in 6-hr bins from 2012 July 22 - December 24 detailing the
pre, post, and three main episodes (cf.,~Figure~\ref{figure-1}, middle),
sub-divided into $\sim8-10$ day-long flares and corresponding delays (asterisks
mark outlying sharp features; see text). Each panel spans 55-days, with adjacent
panels overlapping by 5-days on each side. Horizontal dashed lines indicate the
baseline flux during the flaring interval (Figure~\ref{figure-1}).}
\end{figure*}

In order to study the flaring activity in detail, we defined a 265-day interval 
starting at $T=0$ days and generated 1-day and 6-hr binned lightcurves. The 
\Fermi\ ToO observations also allowed us to produce a $\sim$1.6-hr 
orbit-by-orbit binned lightcurve for the sub-interval covering the first delayed 
flare from 2012 September 24 - October 1 ($T=94-101$ days). To search for any 
possible spectral changes, we initially computed the 1-day binned lightcurve 
with the photon index free in the fit. For the 108 points with the greatest 
significance (test statistic\footnote{The source significance is equivalent to 
$\sim\sqrt{TS}$, assuming one degree of freedom \citep{mat96}.}, 
$TS\geq25$), we found all but four points within $2\sigma$ of the weighted 
average value of $2.31\pm0.02$, which in turn is consistent with the 3.9-yr 
average. We thus regenerated the 1-day (Figure~\ref{figure-1}, middle), the 
1.6-hr orbit (Figure~\ref{figure-1}, bottom), and the 6-hr binned lightcurves 
(Figure~\ref{figure-2}) with $\Gamma=2.3$ fixed.

\section{Results}

\subsection{Time Lag}

\begin{figure}
\epsscale{1.15}
\plotone{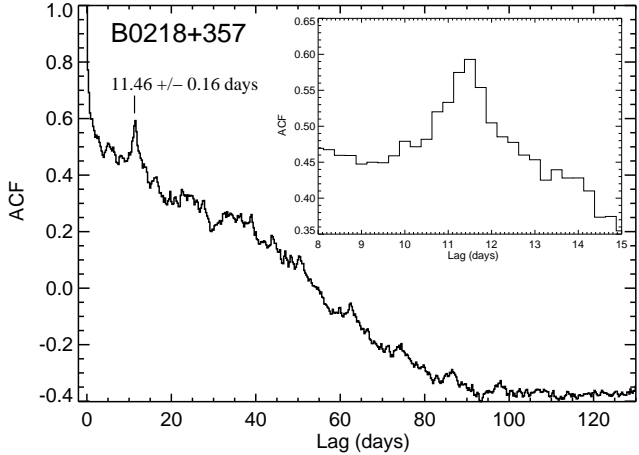}
\figcaption[f3.eps]{\label{figure-3}
Auto-correlation function computed for the 6-hr binned LAT lightcurve of the
265-day flaring interval. The inset zooms in around the best-fit indicated lag
peak.} \end{figure}

The B0218+357 \gray\ lightcurve appears quite complex with many peaks and 
valleys over the $\sim4$ months from $T\sim60-180$ days (Figure~\ref{figure-2}) 
when the source was most active.  To search for a time lag, we computed the 
auto-correlation function (ACF) for the 6-hr binned lightcurve up to lag values 
of half of the total defined 265-day flaring interval. This evenly sampled 
lightcurve consisted of 1057 measurements with three missing data points due to 
exposure gaps. The ACF was therefore computed both by a standard (after 
interpolating the three missing points) and a discrete routine \citep{ede88}. 
The two procedures gave almost identical results and the ACF is shown in 
Figure~\ref{figure-3}. A single prominent correlation peak is apparent between 
the time lag range of $11-12$ days. The peak's significance is $9\sigma$ with 
respect to the measurement noise and comparing it to the height above the ACF 
``background.'' Fitting a Gaussian function to this peak, we estimated a 
best-fit value, $\dtg=11.46\pm0.16$ days ($1\sigma$). Uncertainties were 
estimated by a model independent Monte Carlo method \citep{pet98} accounting for 
the effects of measurement noise and data sampling. The time lag does not match 
any known period observed with the LAT \citep{onorbit,cor12}. Because the \gray\ 
flaring was so pronounced especially from $T\sim84-155$ days, and appears to be 
broadly divided into three $\sim8-10$ day long flare/delay sequences 
(Section~3.2), this could induce other smaller enhancements in the ACF over the 
studied interval.

As a cross-check of the lag derived from the full flaring interval \gray\ data, 
discrete ACFs were computed for two segments from $T=0-110$ and $T=110-265$ 
days. The lags obtained from Gaussian fits to the peaks were, 
$\dtg=11.52\pm0.31$ and $\dtg=11.38\pm0.28$ days, respectively, confirming the 
delay value and small uncertainty for the full interval, thus indicating that we 
obtained a robust measurement with the LAT. The small uncertainty in $\dtg$ is 
competitive with the best determined radio measurements for B0218+357 although 
the former is marginally larger by $\dtg-\dtr=1.0\pm0.3$ and $1.4\pm0.8$ days 
($1\sigma$) than the \citet{big99} and \citet{coh00} values, respectively, but 
consistent with the \citet{eul11} values. If the radio/\gray\ delays are 
intrinsically different due to an offset between the respective emitting 
regions, the implied offset in a singular isothermal sphere lens model is 
$\sim70$ pc (projected) for a $\sim10\%$ difference in the time delay. This 
seems extreme considering such offsets are on average $\sim7$ pc in other blazar 
jets \citep[e.g.,][]{pus10}, and may rather suggest the uncertainty in the radio 
delay was underestimated (Section~1).

\subsection{Flare Timescales}

Utilizing the $\gamma$-ray delay measurement, we can broadly identify three sets 
of flare/delay episodes in the 6-hr binned LAT lightcurve of B0218+357 
(Figure~\ref{figure-2}). The pre-flare times were what triggered the initial 
excitement in late 2012 August and are now detailed as a 6-hr flare at $T\sim50$ 
days (with a corresponding delayed signal 11.5 days later) and a doublet of 
$6-12$-hr flares 1 day apart beginning at $T\sim65$ day. In the doublet, only 
the first flare showed a clear delayed flare 11.5 days later while the second 
shows no similar corresponding delayed (or 11.5-day prior) feature; microlensing 
(see below) or a relatively large variation in the magnification ratio are 
possible explanations.

The first bright \gray\ sequence began at $T=84$ day with the best-defined 
flaring structure with observed fluxes, $\sim(2-5)\times10^{-6}$ \phflux\ over 
eight consecutive 6-hr bins, followed by a sharp drop and subsequent rise in one 
day. The \Fermi\ ToO observation began 10~days later and the anticipated delayed 
emission mirrored the initial flare with the rise and peak separated by 1 day 
and all features well-matched 11.5 days later. We broadly identified two 
subsequent (2nd and 3rd) $\sim8-10$ day duration \gray\ flaring sequences, but 
these were more difficult to disentangle because of superposing flares in the 
integrated lightcurves. The post-flare intervals showed lower fluxes, comparable 
to the pre-flare emission states.

In Figure~\ref{figure-2}, the observed variability timescales (doubling and 
halving), $t_{\rm var}$, during the first and subsequent two flaring episodes 
are securely less than the 6-hr binning. Doubling timescales as short as 
2-orbits ($\sim3$ hrs) are further suggested in the orbit-by-orbit binned 
lightcurve from the \Fermi\ ToO pointing of the first delayed flare 
(Figure~\ref{figure-1}, bottom). Such timescales are amongst the fastest 
well-constrained \gray\ variability in a blazar observed with the LAT 
\citep{tav10,abd11} and constrain the \gray\ emission region diameter, 
$d\leq\,2c\,t_{\rm var}/\,(1+z)\leq6\times10^{14}$~cm, modulo the unknown 
Doppler beaming factor. Assuming an $h=H_{\rm 
0}/(100\,{\rm~km\,s}^{-1}\,{\rm~Mpc}^{-1})=0.71$ ($\Omega_{\rm 
M}=0.27,\,\Omega_{\Lambda}=0.73$) cosmology, this translates to an angular 
diameter $\approx30$ nano-arcseconds, $\sim10^{4}\times$ smaller than the best 
radio size constraint \citep{mit07}. Microlensing is thus an important factor in 
interpreting our \gray\ results because the smaller the structures, the larger 
the expected variability of magnification.

\subsection{Flux and Magnification Ratios}

\begin{figure*}
\epsscale{1.15}
\vspace*{1cm}
\plotone{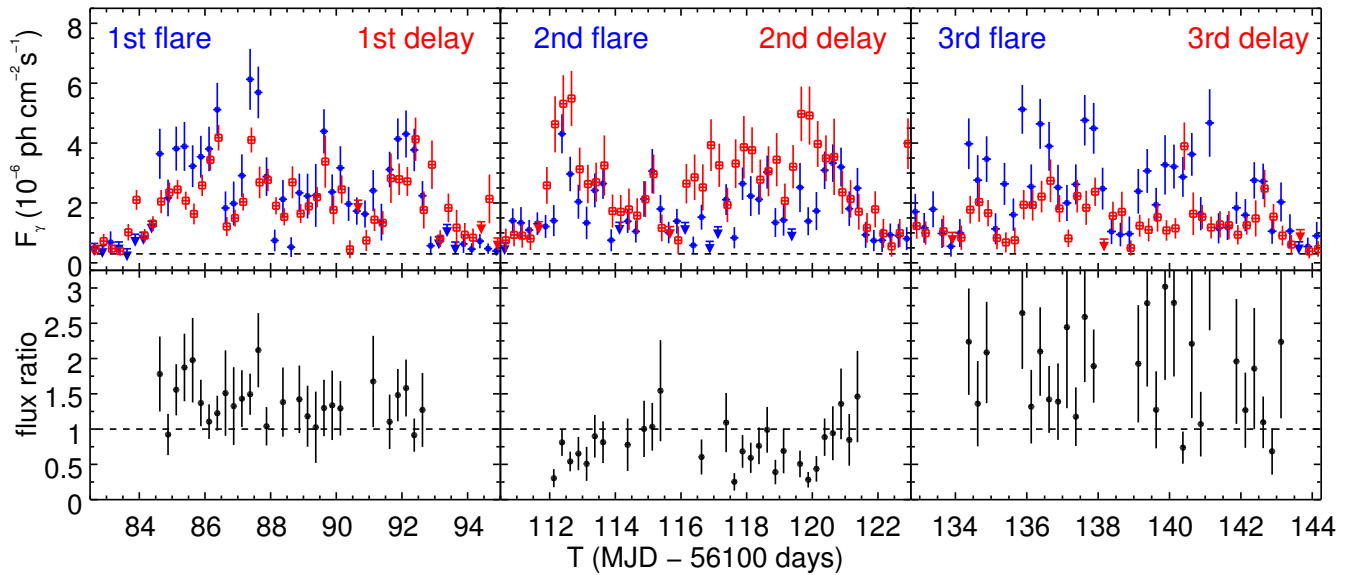}
\figcaption[f4.eps]{\label{figure-4}
Top panels show the 6-hr binned lightcurves (Figure~\ref{figure-2}) for the
three flares (filled blue) and delayed emission shifted by $-11.46$ days (open
red). Bottom panels show individual observed flux ratios (dashed line drawn at
ratio $=1$ for reference) in the corresponding upper panels; error bars are
symmetric and the third panel was cropped in order to display a common range.}
\end{figure*}

Adopting the \gray\ delay, we compared the 6-hr binned lightcurves for the three 
main flaring episodes with the observations shifted by $-11.46$ days and 
computed the observed ratios between corresponding flux pairs, retaining only 
ratio values $\geq 2\times$ their uncertainties (Figure~\ref{figure-4}). The 
first sequence appears to show the clearest correspondence between features in 
the two lightcurves, with only minor deviations about the weighted average flux 
ratio $1.3\pm0.1$. By subtracting a baseline, $F_{\gamma}=0.3\times10^{-6}$ 
\phflux\ (the minimum observed flux during the overall flaring interval), we can 
further estimate a corresponding magnification ratio in \grays\ $\approx1.3$, 
consistent with the flux ratio. The average ratios for this first sequence seem 
to imply the brighter A image led the B image in \grays, as observed in the 
radio. More conservatively however, given the large uncertainties in the 
individual measurements, the flux ratios appear consistent with unity. Moreover, 
for the subsequent 2nd and 3rd sequences, the correspondences between the flare 
and delayed emissions were less clear. Sharp and more scattered changes in the 
paired flux ratios were apparent, including values $<1$ (which would imply a 
fainter leading A image). We interpret this as an artifact due to contamination 
from superposing flares after the source has already entered a very active 
phase. This confusion in the integrated lightcurves prevents us from reliably 
determining magnification ratios, and how variable this quantity may have been.

The flux ratio measured in \grays\ is smaller than in the radio. \citet{big99} 
found a small, but statistically significant frequency dependence in the flux 
ratios, $3.57\pm0.01$ (8 GHz) and $3.73\pm0.01$ (15 GHz), while \citet{coh00} 
found similar values but with larger uncertainty, $3.2^{+0.3}_{-0.4}$ (8 GHz) 
and $4.3^{+0.5}_{-0.8}$ (15 GHz). Frequency dependence in the flux ratios of the 
two radio images and their observed substructures could be possibly due to 
free-free absorption and scattering from a molecular cloud in the lens galaxy 
\citep{mit07}. We note that the radio and \gray\ observations are not 
simultaneous and magnification ratios could be variable with time. Further 
complicating such comparisons are open questions in blazar jet studies, i.e., 
the radio and \gray\ emitting regions need not coincide, with the latter likely 
more compact (Section~3.2), and whether successive \gray\ flares originated in a 
single emission zone or from separate relativistically moving dissipation 
regions along the jet. Excursions could also be due to intrinsic changes in the 
magnification ratios or microlensing from the relative motion of the source seen 
through a clumpy lensing galaxy. Indeed, microlensing in the context of 
extremely compact \gray\ emission zones \citep{tor03} could explain the single 
6-hr flare points that do not have corresponding lags (marked with asterisks in 
Figure~\ref{figure-2}), although fast superposed flares are also a possibility. 
Note that in optical and infrared observations, the B image appears brighter 
than the A image, i.e., reversed from the radio situation, and this is likely 
due to a combination of extinction of the A image and microlensing 
\citep{fal99,jac00}.

\section{Discussion and Conclusions}

Our detection of a gravitational lens time delay, $\dtg=11.46\pm0.16$ days, in 
the LAT observations of blazar B0218+357 has some interesting potential 
implications for future \gray\ studies. Foremost, the LAT detection of a \gray\ 
gravitational lens flaring event in B0218+357 suggests that such a measurement 
is possible in other blazars. In particular, gravitational lenses found in 
surveys of flat-spectrum radio sources \citep{bro03,win00} comprise a relevant 
sample as these form the basis of candidate \gray\ blazar catalogs 
\citep[e.g.,][]{hea07}. There are $\sim20$ gravitational lenses from these 
surveys out of $>10^{4}$ radio sources studied with $\simgt30$ mJy at 8 GHz and 
so far, the two radio brightest are detected \gray\ sources, PKS1830--211 
(below) and B0218+357 \citep[out of $\sim10^{3}$ known \gray\ blazars;][]{2fgl}. 
The other fainter lensed systems are typically less variable at radio 
frequencies, making delay measurements difficult \citep[e.g.,][]{jac07,eul11} 
and while they are not yet reported \gray\ sources, the all-sky monitoring of 
\Fermi-LAT will allow the detection of short-timescale flaring \gray\ activity 
in which to attempt delay measurements. Importantly, \gray\ measurements 
constrain lens parameters free of propagation effects like scintillation 
\citep{hee84,lov08} that can hamper radio delay attempts \citep{win04}, although 
microlensing may be an important limiting factor because \gray\ emitting regions 
are expected to be more compact than in the radio.

The case of B0218+357 appears to be the first clear case of a \gray\ detected 
gravitational lens time delay for any astrophysical system. Previously, \gray\ 
flaring from the gravitationally lensed $z=2.507$ blazar PKS1830--211 was 
detected with the \Fermi-LAT \citep{cip10} with a claimed delay, 
$\dtg=27.1\pm0.6$ days \citep{bar11} consistent with the radio measurement, 
$\dtr=26^{+4}_{-5}$ days \citep{lov98}. Subsequent analysis of more LAT data, 
including several prominent flares, did not confirm the \gray\ delay 
\citep{abd13}. If the \gray\ delay in PKS1830--211 is assumed to be the same as 
the radio-measured delay, the non-detection of delayed \gray\ flares implies a 
magnification ratio in \grays\ much larger ($\simgt6$) than that observed in the 
radio \citep[$1.52\pm0.05$;][]{lov98}, thus opposite of what we observed in 
B0218+357. With only two examples studied, no trend is clear. However, if 
microlensing effects can be disentangled (and in fact utilized as additional 
constraints on the emitting region source size), magnification ratios in radio 
and \gray\ arising from spatially distinct emission regions may be utilized as a 
probe of differing multi-frequency jet structures \citep[see][]{mar13}.

A time delay due to gravitational lensing of a background source by a foreground 
object can constrain Hubble's parameter \citep{ref64}. The original lens model 
for B0218+357 \citep{big99} predicted a delay, $\dt=7.2^{+1.3}_{-2.0}\,h^{-1}$ 
days ($95\%$ confidence). Utilizing improved localization of the lensing galaxy, 
the delay model uncertainty was reduced to $6.0\%$ \citep{yor05}, thus deriving 
$h=0.70\pm0.05$, assuming the often quoted \citet{big99} measured radio delay 
(cf., Section~3.1). Adopting the York model for our independent \gray\ measured 
delay results in $h=0.64\pm0.04$, where this quoted uncertainty is due only to 
the time delay estimate and the statistical uncertainty in the mass model. 
Systematic errors in the modeling, and additional uncertainty due to 
line-of-sight structures \citep[e.g.,][]{suy12} will likely significantly 
increase this. Nevertheless, it is interesting that the LAT time delay brings 
the estimated value of Hubble's constant down, towards the low end of modern 
measurements \citep[e.g.,][]{pla13}. An underdense environment would require 
this inferred $h$ value to increase; including external lensing effects in 
future cosmographic analyses might be important in this system. Moreover, since 
the radio and \gray\ emission regions are likely not co-spatial, the assumed 
radio-derived time-delay function values may be inaccurate. A fully 
self-consistent joint modeling of the radio and \gray\ source is needed to 
resolve this. If the LAT can measure a lag in the \gray\ lightcurve of one of 
the previously known systems with wider separation or in a new example (below), 
this can give independent \gray\ based constraints on Hubble's constant.

One exciting result would be the detection of a lens delay in a flaring \gray\ 
source that is not yet identified as a gravitationally lensed system at radio 
wavelengths or otherwise. These could possibly be lensed image pairs with 
flat-spectrum radio sources at smaller separations than in the 0.2\arcsec\ 
resolution of VLA surveys (references above). Similar radio lens surveys in the 
southern hemisphere are not yet as complete \citep[e.g.,][]{pro99}, so a \gray\ 
delay signature in their LAT lightcurves could betray the presence of a 
previously unknown lens system. Such a strategy has been proposed for future 
wide-field optical surveys \citep{pin05}, and the discovery potential of the LAT 
in \grays\ should now be recognized. Furthermore, with the different flux ratios 
at radio and \gray\ wavelengths, and possible variability of the ratio, some 
sources could be bright in \grays\ and less conspicuous at radio. Such potential 
gravitational lenses could be hidden in plain sight within the radio catalogs 
used for blazar associations in LAT catalogs, or could be amongst the currently 
unidentified \gray\ sources \citep{tor02}.

\acknowledgments

The \Fermi-LAT Collaboration acknowledges support from a number of agencies and 
institutes for both development and the operation of the LAT as well as 
scientific data analysis. These include NASA and DOE in the United States, 
CEA/Irfu and IN2P3/CNRS in France, ASI and INFN in Italy, MEXT, KEK, and JAXA in 
Japan, and the K.A.~Wallenberg Foundation, the Swedish Research Council and the 
National Space Board in Sweden. Additional support from INAF in Italy and CNES 
in France for science analysis during the operations phase is also gratefully 
acknowledged. 
C.C.C.~was supported at NRL in part by NASA DPR S-15633-Y.

{\it Facilities:} \facility{Fermi}

{}

\end{document}